# High stability 2D electron gases formed in $Si_3N_4$/Al//$KTaO_3$ heterostructures: synthesis and in-depth interfacial characterization

E. A. Martínez[a,b], A. M. Lucero[c,d,e], E. D. Cantero[c,d], N. Biškup[a,f], A. Orte[a], E. A. Sánchez[c,d], M. Romera[a], N. M. Nemes[a,g], J. L. Martínez[g], M. Varela[a,f], O. Grizzi[c,d] and F. Y. Bruno[a*]

[a]GFMC, Departamento de Física de Materiales, Universidad Complutense de Madrid, Madrid 28040, Spain

[b]Dipartimento di Scienze Fisiche e Chimiche, Università degli Studi dell'Aquila, L'Aquila 67100, Italy

[c]Centro Atómico Bariloche, Comisión Nacional de Energía Atómica (CNEA), Bariloche 8400, Argentina

[d]Instituto de Nanociencia y Nanotecnología, Consejo Nacional de Investigaciones Científicas y Técnicas (CONICET), Bariloche 8400, Argentina

[e]Universidad Nacional de Cuyo, Facultad de Ciencias Exactas y Naturales, Mendoza 5500, Argentina

[f]Instituto Pluridisciplinar, Universidad Comlutense de Madrid. Madrid 28040, Spain

[g]Instituto de Ciencia de Materiales de Madrid, ICMM-CSIC, Cantoblanco, Madrid 28049, Spain

*Corresponding author: fybruno@ucm.es

**Abstract**

The two-dimensional electron gas (2DEG) found in $KTaO_3$-based interfaces has garnered attention due to its remarkable electronic properties. In this study, we investigated the conducting system embedded at the $Si_3N_4$/Al//KTO(110) heterostructure. We demonstrate that the Al/KTO interface supports a conducting system, with the $Si_3N_4$ passivation layer acting as a barrier to oxygen diffusion, enabling ex-situ characterization. Our findings reveal that the mobility and carrier density of the system can be tuned by varying the Al layer thickness. Using scanning transmission electron microscopy, electron energy-loss spectroscopy, X-ray photoemission spectroscopy, and time-of-flight secondary ion mass spectrometry, we characterized the structural and chemical composition of the interface. We found that the Al layer fully oxidizes into $AlO_x$, drawing oxygen from the $KTaO_3$ substrate. The oxygen depletion zone extends 3–5 nm into the substrate and correlates to the Al thickness. Heterostructures with thicker Al layers exhibit higher carrier densities but lower mobilities, likely due to interactions with the oxygen vacancies that act as scattering centers. These findings highlight the importance of considering the effect and extent of the oxygen depletion zone when designing and modeling two-dimensional electron systems in complex oxides.

1. Introduction

The bulk physical properties of transition metal oxides (TMOs) are primarily governed by the strongly correlated *d* electrons. In these materials, the intricate interplay between lattice, charge, spin, and orbital degrees of freedom results in a plethora of phenomena, including superconductivity, magnetism, ferroelectricity, and charge and orbital ordering [1]. In 2004, a groundbreaking discovery revealed the formation of a high-mobility two-dimensional electron gas (2DEG) at the interface between two band insulating oxides, $LaAlO_3$ (LAO) and $SrTiO_3$ (STO) [2]. This finding, together with the diverse electronic phases observed in complex oxides, has spurred extensive efforts to harness these properties in functional devices, thereby initiating the field of oxide electronics [3,4]. Significant research efforts within the oxide electronics community have been devoted to understanding the underlying mechanisms, discerning variations in physical properties, and exploring potential applications of this 2DEG. Numerous intriguing phenomena associated with this oxide-based 2DEG have since been observed, including superconductivity [5], a unique magnetic response [6,7], and an unconventional Rashba effect [8–10]. It was later discovered that KTaO$_3$ (KTO)-based heterostructures also host a 2DEG with remarkable properties [11–13]. KTO and STO share several properties: both are band insulators with about 3 eV gap, are quantum paraelectric, and have the cubic perovskite structure at room temperature. However, a significant difference lies in their spin-orbit coupling (SOC) strength. KTO has a much larger SOC, 0.4 eV for the Ta *5d* electrons, compared to approximately 0.02 eV for the Ti *3d* electrons in STO. The high SOC of KTO, related to an expected large Rashba effect in the 2DEG, prompted researchers to look for applications of this KTO-based 2DEG system in spintronics [14,15]. The unanticipated recent discovery of superconductivity in (111) and (110) KTO based 2DEGs, with superconducting critical temperature ($T_c$) values up to five times higher than those reported for STO-based 2DEGs have brought the study of the electronic properties of KTO based 2DEGs back into the forefront of material science [16–18].

In order to stabilize a 2DEG in KTO several strategies can be used. It has been shown that depositing EuO, LaAlO$_3$, LaTiO$_3$ layers, among others, on top of a KTO substrate results in the creation of a 2DEG[12,17,18]. The properties of the conducting system are related to the crystallographic orientation of the substrate, the carrier density and the disorder near the interface; the latter can be altered as a function of growth parameters[18–21]. A simpler route to generate the 2DEG is the growth of thin Al layer on top of a KTO single crystal. This method originally reported for STO[22], has been proven to also work for KTO.[14,23]. The growth of this metallic layer results in a disordered and oxidized Al layer, where the oxygen is pumped from the KTO creating oxygen vacancies (V$_O$'s) near the substrate surface[22–24]. One key advantage of the method is that high quality angle resolved photoemission spectroscopy measurements of the 2DEG electronic structure can be done after Al deposition, thus providing a method for direct comparison with transport experiments[22,23,25]. While these experiments provide evidence for the existence of V$_O$'s in Al/KTO and Al/STO interfaces, measurements of their spatial spread are scarce [26,27]. The distribution of oxygen vacancies near the surface, which act as electron dopants, can significantly impact the electronic properties of the system [28]. A particularly important case arises when the region of electron confinement is similar in size to the extent of V$_O$'s. In this scenario, the vacancies can serve as scattering centers for charge carriers. This is highly relevant because disorder and mobility have been identified as the key parameters controlling the superconducting transition temperature [21]. Additionally, the presence of localized states associated to the V$_O$'s can significantly modify the confinement potential. Many models assume that this potential is created by an excess of electrons populating the conduction bands at the TMO surface/interface[29].

In this work, we present a robust growth protocol for generating KTO-based 2DEGs using Si$_3$N$_4$/Al//KTO heterostructures fabricated by magnetron sputter deposition. Amorphous Al layers, nominally 1 nm and 2.5 nm thick, are grown to create the 2DEG at the Al/KTO interface, promoting the generation of V$_O$'s on the substrate surface and subsequent Al oxidation. The Si$_3$N$_4$ layer, commonly used as a passivation layer in microelectronics, acts as a capping that serves as an insulating barrier against oxygen diffusion [30,31]. We then conduct *ex-situ* electrical and structural characterizations of the Si$_3$N$_4$/Al//KTO(110) heterostructures, given the interesting highly anisotropic features that the KTO(110)-2DEGs present [17,23,32,33]. Through magneto-transport experiments, we demonstrate that the 2DEG is effectively stabilized at the Al/KTO interface and that Si$_3$N$_4$ serves as an efficient capping layer, making the 2DEG non-volatile. We observe that the 2DEG carrier density and mobility can be altered by varying the Al layer thickness, and we study the temperature dependence of these properties. For structural characterization, we introduce time-of-flight secondary ion mass spectrometry (TOF-SIMS) as a novel approach to study the heterostructure composition, leveraging the high chemical sensitivity of the technique. Several studies have applied TOF-SIMS to perovskite structures [34–37] but none have focused on studying 2DEGs in perovskite oxides in combination with other techniques, as presented in this work. Using TOF-SIMS, we estimate the extent of V$_O$'s inferred from the behavior of specific chemical species. Combining this technique with scanning transmission electron microscopy (STEM), electron energy-loss spectroscopy (EELS), and X-ray photoelectron spectroscopy (XPS), we obtain a detailed compositional description along the 2DEG confinement direction [38,39]. Our findings suggest a correlation between the extent of oxygen depletion region and the 2DEG transport properties.

## 2. Experimental methods

The $Si_3N_4/Al//KTaO_3$ heterostructures were created by growing amorphous Al and $Si_3N_4$ layers on top of KTO single crystals by magnetron sputter deposition [40]. The KTO substrates undergo annealing at 500 °C for 30 min under a pressure better than $10^{-7}$ mbar. Then the Al layer is deposited by DC magnetron sputtering under an Ar atmosphere of $5.4 \times 10^{-3}$ mbar. Subsequently, the $Si_3N_4$ layer is deposited by RF magnetron sputtering at the same Ar pressure. After removing the samples from the chamber, electric contacts were created by Al-wire ultrasonic bonding in van der Pauw configuration [41], and the magneto-transport characterization was performed in a physical properties measurement system (PPMS). In order to perform structural investigations of the samples, STEM was employed to provide a nanometric-scale characterization of the heterostructure. Through the integration of STEM with EELS (STEM-EELS), we have obtained a detailed characterization of the chemical composition of the distinct layers and their thickness within the $Si_3N_4/Al//KTaO_3$ samples. To complement these findings, we employed XPS while performing an $Ar^+$ depth profiling of the sample, mainly intended to study the oxidation states of the Al across the heterostructure. Finally, we employed TOF-SIMS, in its depth profiling operation mode, to complement our findings from STEM-EELS and XPS, which allows us to depict a precise scheme of the heterostructure conformation. Further information about the growth protocol and each of the techniques employed can be found in the Methods section.

## 3. Results

We fabricated several $Si_3N_4/Al//KTO$ heterostructures for the (001), (110) and (111) faces, which show sheet resistances ($R_s$) in the 10-30 kΩ sq$^{-1}$ range at RT with an Al thickness of nominally 1 and 2.5 nm. An insulating $Si_3N_4$ passivation layer with nominal thickness between 5 and 20 nm is used to prevent oxygen diffusion when the sample is exposed to air. Although the method of stabilizing a 2DEG by depositing a thin Al layer has been previously used in KTO and other oxides, we verified that is indeed effective in our heterostructures. Replacing the KTO substrate with $LaAlO_3$ (LAO) results in insulating ($dR_s/dT<0$) heterostructures, proving that the deposited Al layer has a negligible conductivity. Eliminating the Al layer, namely, opting to deposit $Si_3N_4$ on the KTO surface, also yields insulating samples. If the $Si_3N_4$ capping is not deposited, the resulting Al//KTO samples also turned out to be insulating due to the Al/KTO oxidation in atmospheric conditions. Notably, the $Si_3N_4/Al//KTO$ heterostructures grown at RT, without annealing of the substrate in vacuum also turned-out insulating. From these series of experiments, we confirmed the formation of a 2DEG when depositing a thin Al layer on the clean surface of KTO. The efficient redox reaction at the Al/KTO interface results in localized positively charged $V_O$'s and confined mobile electrons that screen these charges [14,22,23,25,42,43]. Therefore, the interface is formed by $AlO_x/KTO$, where x represents a fraction of O transferred to the Al film. However since the material grown is Al, we will maintain the Al/KTO notation along the manuscript. It has been shown in photoemission experiments that the presence of states at the Fermi level, a signature of a 2DEG, can be detected with Al layers as thin as 0.3 nm. In the present study, to measure low resistance values in transport experiments, an Al thickness of at least 1 nm is necessary. We attribute this discrepancy to the inhomogeneous growth of the Al layer [22,43,44]. Photoemission experiments can detect the electronic states for very thin Al thickness as the insulating parts lack spectral weight at the Fermi level. In contrast, during transport experiments, a percolation path is necessary to measure their high conductivities. In Table 1 below, we show the samples that we have measured in this study along with their $R_s$ values measured in a four-probe station in atmospheric conditions (see Methods). Additional samples are summarized in Table S1 of Supplementary Information.

**Table 1:** Samples prepared for magneto-transport characterization using a PPMS setup and their sheet resistance, $R_s$, measured at RT with a four-probe station. The samples were contacted in the van der Pauw configuration by ultrasonic bonding with Al-wire. The "†" and "‡" superscripts denote samples that were grown within the same batch. The "∞" denotes that the resistances resulted too high, exceeding the measurement limit for the power supply used. Superscripts A and B indicate different KTO substrate providers.

| Sample name | Heterostructure | $R_s$(kΩ sq$^{-1}$) at RT |
|---|---|---|
| Al-KTO#1[†] | Si$_3$N$_4$/Al(2.5 nm)//KTO(110)[A] | 11 |
| Al-KTO#2[†] | Si$_3$N$_4$/Al(2.5 nm)//KTO(110)[B] | 16 |
| Al-KTO#3[‡] | Si$_3$N$_4$/Al(1 nm)//KTO(110) | 12 |
| Al-LAO#1[†] | Si$_3$N$_4$/Al(2.5 nm)//LAO(001) | ∞ |
| Al-LAO#2[‡] | Si$_3$N$_4$/Al(1 nm)//LAO(001) | ∞ |

The temperature dependence of $R_s$ for various heterostructures is shown in Fig. 1 together with the carrier density ($n_s$) and mobility ($\mu$) as deduced from the Hall measurements shown in Section 2 of the Supplementary Information. In the present study, we will focus on the 2DEG stabilized on samples grown on KTO(110), measurements for systems confined in other crystallographic orientations can be found in Section 1 of the Supplementary Information. The metallic ($dR_s/dT>0$) behavior of samples Al-KTO#1, Al-KTO#2, and Al-KTO#3 is demonstrated in Fig. 1(a) consistent with the presence of the 2DEG [17,18,21,45]. To qualitatively assess the metallic nature of each sample, we compute the residual-resistivity ratio (*RRR*), defined as *RRR* = $R_s(T=300\text{ K})/R_s(T=2\text{ K})$, which indicates the overall interfacial quality [46,47]. Notably, Al-KTO#3 exhibits the highest metallicity with a *RRR* of 11.5, broadly surpassing the values of 6.75 and 4.74 for Al-KTO#1 and Al-KTO#2, respectively. This heightened metallic behavior in Al-KTO#3 aligns with its high mobility, and it could be due either to a better interface quality or to a smaller amount of V$_O$'s, factors that could limit scattering or charge carrier trapping. Similar heterostructures grown on LaAlO$_3$ substrates, displayed in Fig. 1 as Al-LAO#1 and Al-LAO#2, are at least 4 (8) orders of magnitude more resistive at RT (10 K). Thus, the effect of non-oxidized Al clusters or the defective Si$_3$N$_4$ layer can be ignored in the analysis of the transport properties of the 2DEG. Samples Al-KTO#1 and Al-KTO#2 exhibit low *T* carrier densities of $n_s \approx 4.5$ and $6.5\times10^{13}$ cm$^{-2}$, associated with mobilities of $\mu \approx 40$ and 60 cm$^{-2}$ V$^{-1}$ s$^{-1}$, respectively. The $n_s$ for these samples increases up to approximately 5.8 and $8.0\times10^{13}$ cm$^{-2}$, at RT, while there is a decline in $\mu$ by an order of magnitude, reaching $\mu \approx 7$ cm$^{-2}$ V$^{-1}$ s$^{-1}$ for both samples. This result was expected, given that both samples were grown simultaneously but in substrates of different providers, indicating that the 2DEG is sensitive to the surface morphology and/or crystal quality. The behavior of sample Al-KTO#3 is markedly different. The $n_s$ for this sample does not follow a monotonic trend, and its low *T* mobility is elevated, as reported in other KTO- and STO-based 2DEGs [17,48–50]. At low *T*, the carrier density remains stable at a value of $n_s \approx 3.8 \times 10^{13}$ cm$^{-2}$ with a notably high mobility of $\mu \approx 165$ cm$^{-2}$ V$^{-1}$ s$^{-1}$. At 10 K, $n_s$ experiences a sudden rise, peaking at approximately $4.9\times10^{13}$ cm$^{-2}$ around 100 K. From 100 K to 300 K, $n_s$ decreases to a value of $\approx 4.5\times10^{13}$ cm$^{-2}$. This fact could be associated with some carrier re-trapping mechanisms at such low densities, as reported in other works [49,51,52]. Simultaneously, the mobility undergoes a sharp reduction above 10 K, settling at $\mu \approx 12$ cm$^{-2}$ V$^{-1}$ s$^{-1}$ at 300 K. The values observed in our samples align with those documented for KTO-based 2DEGs in other orientations [17,18,45].

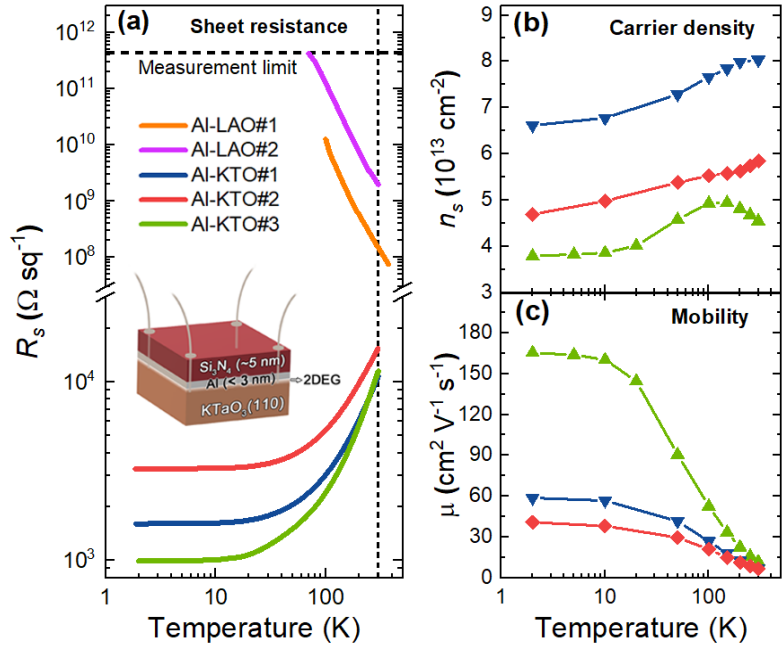

**Figure 1. (a)** Sheet resistance $R_s$ as a function of temperature $T$ for samples Al-KTO#1, Al-KTO#2, Al-KTO#3, Al-LAO#1 and Al-LAO#2. *Inset:* schematics of the ultrasonic bonded Al-KTO samples indicating where the 2DEG is confined. A remarkable difference in the $R_s$ behavior is present by changing the substrate. The Al-KTO samples are conductive, whereas Al-LAO samples are insulating with a $R_s$ of at least four orders of magnitude higher than the values measured for the Al-KTO ones. Sheet carrier density $n_s$ **(b)** and mobility $\mu$ **(c)** as a function of $T$ for the conductive samples Al-KTO#1, Al-KTO#2 and Al-KTO#3. The behavior of $n_s$ is as expected when compared to recent reports. For the lowest $n_s$, the greatest $\mu$ is found at low $T$, which is presumably due to a smaller number of scattering centers for such low carrier densities.

Now we turn our attention to the structural characterization of a $Si_3N_4/Al(2.5\ nm)//KTO(110)$ heterostructure, i. e., a sample with the same nominal Al layer thickness as Al-KTO#1. In Fig. 2(a), a scheme of the unit cell of KTO is shown along with a representation of the (110) surface. In Fig. 2(b) an annular dark field (ADF) image is presented, the contrast between the layers is related to the atomic number of the atoms. The single crystal structure of the substrate is clearly resolved; in contrast, the sputter grown layers are amorphous. We have acquired elemental maps in the area marked in Fig. 2(b) using electron energy-loss spectroscopy (EELS), a high magnification ADF image of the interface and the chemical species detected by EELS are depicted in Fig. 2(c). Layers appear flat and continuous, and the sequence confirms the nominal composition based on the heterostructure growth. The distribution of N and Si signals overlap, conforming the $Si_3N_4$ passivation layer, followed by the Al layer, and finally Ta and K signals appear within the substrate area. Some degree of roughness is observed in the Al interfaces, probably related to a 3D-like growth. This aligns with our previous observation of non-conductivity in the samples with Al thickness in the 0.1-0.3 nm range, likely associated with layer discontinuity at such ultrathin limit and, hence, lack of percolation in the 2DEG. Regarding the O signal, a high intensity is detected on the surface, associated with the formation of $SiO_x$ and adsorption of contaminants. When moving across the capping layer, a dip is observed in the O signal due to the presence of the $Si_3N_4$, which acts as a barrier for O diffusion. The O signal in the EELS map then exhibits a sharp rise upon entering the Al layer, pointing to oxidized Al. These findings are further supported by the XPS spectra shown in Fig. 2(d) acquired while measuring a depth profiling of the heterostructure. The Al $2s$ levels in XPS shift by approximately 2 eV from the $Al^0$ $2s$ binding energy throughout the layer, denoting the presence of $Al^{3+}$ which confirms that the entire Al layer is oxidized. We determined a thickness of (2.7±0.4) nm and (1±0.2) nm for samples with nominal Al layer of 2.5 and 1 nm, consistent with the calibration of the source. Details of the thickness determination along with additional STEM-EELS imaging and a XPS depth profiling following the O $1s$ core level can be found in Sections 4 and 5 of Supplementary Information.

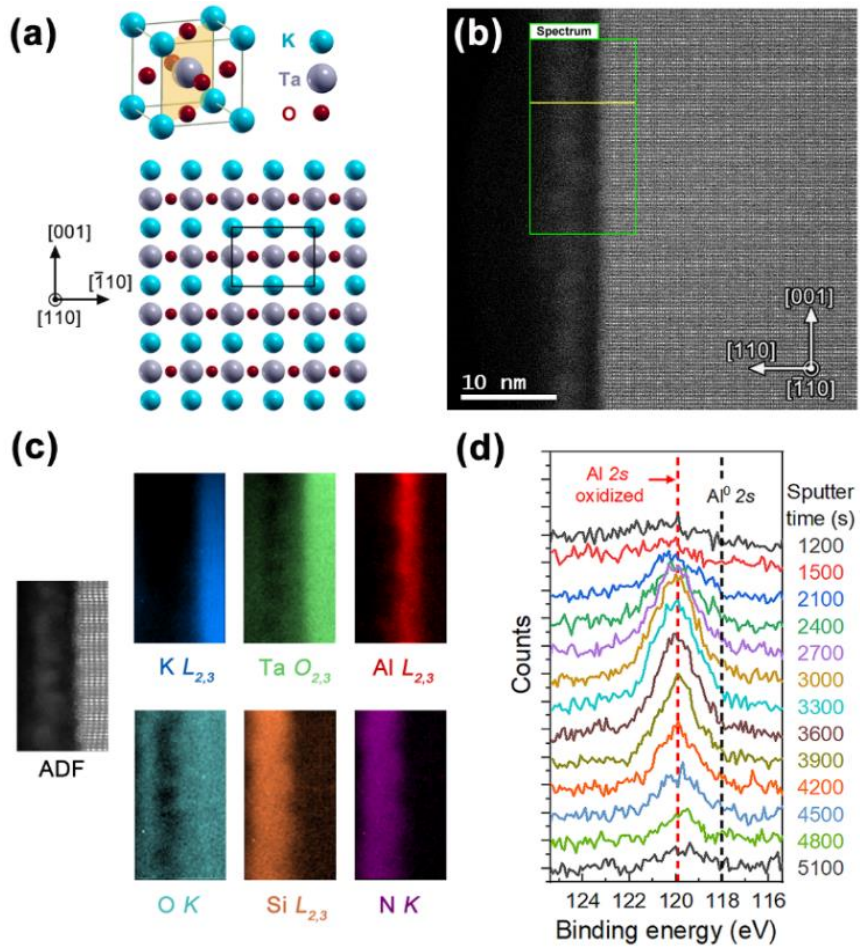

**Figure 2.** STEM-EELS and XPS analysis of the Si$_3$N$_4$/Al(2.5 nm)//KTO(110) interface. **(a)** Schematics showing the bulk unit cell of KTaO$_3$, and top-view of the 2D unit cell for the KTaO$_3$(110) surface. The crystallographic axes are also indicated. **(b)** ADF image showcasing the interface, viewed down the <110> projection. The green rectangle marks a region where an EEL spectrum image was acquired. **(c)** EELS-derived chemical contrast images for the region specified in (b), displaying the spatial distribution of K, Ta, Al, O, Si, N signals under the edges tagged in the figure. **(d)** XPS spectra of the Al *2s* core level during a depth profiling. The shift of the peak towards a higher binding energy compared to Al$^0$ confirms that the Al layer is fully oxidized.

Unlike the STEM-EELS measurements, where elements are identified by their characteristic electronic transitions, the TOF-SIMS technique identifies chemical entities based on ionized atoms and clusters that are ejected from the surface during the sputtering process. The combination of Time-Of-Flight sensitivity in chemical detection and the vast array of ionized atoms, isotopes, and clusters that can be sputtered from the surface, offers a unique perspective into the chemical features across the interface [53]. Figs. 3(a) and (b) display TOF-SIMS depth profiles for positive and negative ions, respectively, for a Si$_3$N$_4$/Al(2.5 nm)//KTO(110) heterostructure. Importantly, the fact that both negative and positive ions display similar depth dependence, and are independent of the sputter energy, underscores this as an intrinsic trend, not a spurious one due to electron transfer in different matrices or substrates. In Fig. 3(a), Al$^+$ and Al$_2^+$ profiles highlight a similar initiation depth. However, Al$^+$ stretches about 1-2 nm further, with a slightly shifted peak, indicative of reimplantation effects or potential Al diffusion. Since reimplanted elements disperse more within subsequent layers, their likelihood of sputtering as clusters diminishes in this zone. We therefore assume that clusters, like Al$_2^+$ and Si$_3$N$_3^-$, offer better delineation of layer ending. On the other hand, the layer initiation seems to be minimally impacted by reimplantation. For this reason, a criterion of peak half maximum to define the beginning of a new layer was considered, the substrate region was characterized by the $^{41}$K$^+$ signal. The layer extension is indicated by the background color in Fig 3 and Fig 4, the overlapping regions are related with the reimplantation process and roughness. Interestingly, the Si$^+$ and O$^+$ signals show an increase at the surface due to SiO$_x$ formation, as revealed by the XPS depth profiling shown in Section 5 of the Supplementary Information. The O$^+$ signal dips within the capping, and increases within the Al layer, being consistent with the Al layer oxidation as confirmed by XPS in Fig. 1(d). This signal finally stabilizes ~5 nm deeper than the substrate interface, suggesting that O migrates from the substrate into the Al layer. For the negative ion profiles in Fig. 3(b), the behavior of the O$_2^-$ is basically the same as in the positive profiles, reinforcing the evidence of the O migration from the substrate. It is worth noting that the Si$^+$ and Si$_3$N$_3^-$ behavior deviates from the expected, its extension into the Al layer surpasses typical reimplantation or broadening processes, as also happens in the positive profiles with Si$^+$ (Fig. 3(a)). We attribute this to the high roughness and/or 3D like growth of the Al layer as previously suggested by the STEM-EELS data. In the Al region very thin sections or even voids could emerge, thus bringing the capping species distributions nearer to the substrate. In correspondence with this, the width of the Al profile is larger than the expected because it is determined by the height of the Al structures and not by the average (nominal) width (2.5 nm) of the Al layer

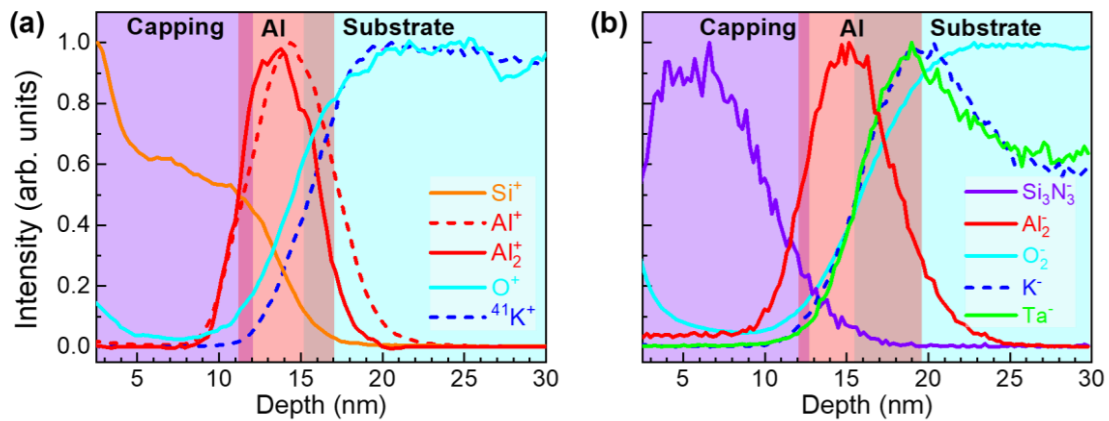

**Figure 3.** Normalized TOF-SIMS profiles for diverse representative species of the Si$_3$N$_4$/Al(2.5 nm)//KTO(110) interface. **(a)** Positive ion profile with sputtering of Cs$^+$ at 500 eV. **(b)** Negative ion profile with sputtering of Cs$^+$ at 250 eV. The background color indicates the extension of the different layers with some overlapping due to reimplantation and topographic effects as described in the text.

As shown in Figure 3, the O$^+$ and O$_2^-$ signal provides some evidence of the oxygen migration from the substrate. We focus our attention now on the determination of the width of the oxygen depletion layer. In figure 4(a) and (b) we show TOF-SIMS profiles for a Si$_3$N$_4$/Al(1 nm)//KTO(110) and a similar heterostructure with no Al layer and hence no expected oxygen diffusion from the KTO, respectively. Close to the substrate surface, the distance between the Ta$^-$ and O$_2^-$ signal maximum is larger for the sample with Al, again pointing in the direction of lack of oxygen close to the interface. To further verify this hypothesis, attention is devoted to the KTaO$_3^-$ and KTaO$^-$ fragments. In Fig. 4(a), we observe that the KTaO$^-$ signal has a peaked structure at the Al/KTO interface and it extends ~3-4 nm into the bulk, in contrast, the KTaO$_3^-$ fragment dips within the same region. This can be interpreted as the migration of O into the Al region, making the KTaO$^-$ signal increase, given that it requires less O to be formed. On the other hand, the KTaO$_3^-$ decreases because of the lack of O. Consistent with this observation, if we track the same profiles for the interface with no Al as in Fig. 4(b), both KTaO$_3^-$ and KTaO$^-$ signal intensities remain equal across the interface, meaning that there is no O-depletion in this scenario. A similar result is observed in a bare KTaO$_3$ substrate as shown in Section 6 of the Supporting Information. Finally, having established the fact that the KTaO$_3^-$ and KTaO$^-$ fragments signals are reasonable indicators of the V$_O$'s extent, we plot in Fig. 4(c) these quantities for Si$_3$N$_4$/Al(x nm)//KTO(110) heterostructures with x = 1 and 2.5 nm. Using our previously defined criteria for thickness estimation in TOF-SIMS, we determine the V$_O$'s region by the depths where the KTaO$_3^-$ and KTaO$^-$ fragments reach half their maxima. From this, we deduce V$_O$'s extents of ~4.5 nm and ~3 nm for the samples with Al nominal thickness of 2.5 and 1 nm, respectively.

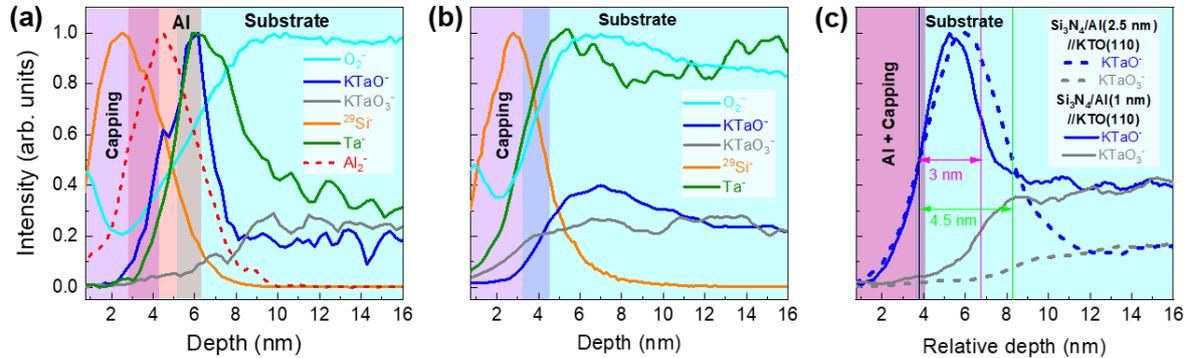

**Figure 4.** Visualization of the O-depleted region within the substrate. **(a)** Depth-dependent negative ion profiles of a Si$_3$N$_4$/Al(1 nm)//KTO(110) interface highlighting representative species from each layer for the Al-KTO#3 sample. KTaO$^-$ and KTaO$_3^-$ profiles are included to shed light on the V$_O$'s extent. **(b)** Analogous measurement as in (a) for the Si$_3$N$_4$/KTO(110) structure. The consistent behavior between KTaO$^-$ and KTaO$_3^-$ fragments suggests that O vacancies arise primarily from Al oxidation. **(c)** Comparison of the KTaO$^-$ cluster extent into the substrate between Al-KTO#1 and Al-KTO#3 samples.

4. **Discussion and conclusions**

In our study of the heterostructures conformation, we have measured the thickness and composition of the layers by combining diverse techniques. The main advantage of TOF-SIMS is that it allows for the measurement of the O-depletion region within the substrate, *i. e.*, the V$_O$'s extent, information hard to get from other techniques. Recalling Figs. 4(b) and (c), we can observe clearly that the V$_O$'s arise from the presence of Al, and that it increases with the Al thickness. In short, for heterostructures with 1 nm (2.5 nm) of Al thickness, we observe an V$_O$'s extent of ~3 nm (4.5 nm). As expected, the thicker the Al layer the thicker the V$_O$'s extent. Some theoretical and experimental investigations in KTO and STO suggest that V$_O$'s can form different defects, associated with specific localized states. Part of the electrons can be trapped in such states, whereas others can populate the conduction band of the TMO [52,54–56]. In such a scenario, the increase in $n_s$ can result from heightened free carrier availability due to the abundance of V$_O$'s. This outcome in transport properties is a relevant link between the electrical and structural characterizations. Indeed, from our results it is known that $n_s$ (Al-KTO#3) < $n_s$ (Al-KTO#1), and consequently, the high mobility of the Al-KTO#3 sample could be associated with fewer scattering centers derived from the shorter V$_O$'s extent in samples with a thinner Al layer. Additionally, free carriers can more effectively screen the scattering centers in low density 2DEGs, especially with the high relative permittivity of KTO at low temperatures[57–59]. An opposite behavior is observed for the Al-KTO#1 sample, where the mobility is reduced, being consistent with the larger V$_O$'s extent observed that can promote more electrons to the 2DEG conduction levels while more scattering centers are introduced. As the temperature increases, we see a gradual rise in the $n_s$ value, likely due to thermal excitation, allowing electrons to cross the energy barrier from donor localized in-gap states to the 2DEG conduction levels. It indicates that our samples present a freeze-out effect as reported elsewhere for these systems [18,45,60]. Given that electron-lattice and electron-phonon interactions become more prominent at high *T*'s, electron re-trapping by the in-gap states becomes more likely [52,61]. Considering our STEM-EELS and TOF-SIMS results, the substantial number of V$_O$'s likely produces a considerable number of in-gap states. The trapped electrons in the localized states would produce a significant modification in the potential confinement profile and the associated band structure (see Section 3 of the Supplementary Information). For the sample Al-KTO#1, the larger amount of V$_O$'s (TOF-SIMS analysis of a Si$_3$N$_4$/Al(2.5 nm)//KTO(110) sample) produces a significant number of trapped electrons along the whole *T* range. On the other hand, for the Al-KTO#3 sample, the smaller amount of V$_O$'s (TOF-SIMS analysis of a Si$_3$N$_4$/Al(1 nm)//KTO(110) sample) probably leads to a competition between de-trapping and re-trapping mechanisms that depend on the *T* range, $n_s$, defect presence, and interface quality [49,51,52]. Thanks to the combination of all the techniques, we can stablish that the larger the amount of Al the deeper the V$_O$'s extent, and the lower the mobility and vice versa. The values discussed here and other quantities reported throughout the work are summarized in the Table 2 below, where the samples are differentiated by the nominal Al thickness.

**Table 2:** Summary of the most relevant quantities for the Si$_3$N$_4$/Al(x)//KTO(110) heterostructures, with nominal Al thickness of x=2.5 nm and x= 1 nm. Here we include the results arising from STEM-EELS, TOF-SIMS and the magneto-transport results for the analogous samples respecting to the Al thickness. Note that the rows explicitly containing the names Al-KTO#N (N=1,2,3) correspond to the data obtained from magneto-transport experiments. The nominal thickness is that estimated prior to the characterizations, while the STEM-EELS and TOF-SIMS thickness are those obtained by each respective technique. Then we show the oxygen vacancies (V$_O$'s) extent, as obtained by TOF-SIMS. The RRR is the residual-resistance ratio that we chose to assess qualitatively how metallic the systems are, defined as RRR ≡ $R_s$(*T*=300 K)/$R_s$(*T*=2 K). Finally, the values for sheet carrier densities and mobilities at low/high temperatures are also included.

| Samples | Si$_3$N$_4$/Al(2.5 nm)//KTO(110) (equivalent to Al-KTO#1 and Al-KTO#2) | | Si$_3$N$_4$/Al(1 nm)//KTO(110) (equivalent to Al-KTO#3) |
|---|---|---|---|
| **TEM-EELS thickness (nm)** | 2.7±0.4 | | 1±0.2 |
| **V$_O$'s extent (nm)** | ∼4.5 | | ∼3 |
| *RRR* | **Al-KTO#1:** 6.75 | **Al-KTO#2:** 4.74 | **Al-KTO#3:** 11.2 |
| $n_s$ at 2/300 K (10$^{13}$ cm$^{-2}$) | **Al-KTO#1:** 4.5 / 5.8 | **Al-KTO#2:** 6.5 / 8.0 | **Al-KTO#3:** 3.8 / 4.5 |
| $\mu$ at 2/300 K (cm$^{-2}$ V$^{-1}$s$^{-1}$) | **Al-KTO#1:** 59 / 7 | **Al-KTO#2:** 41 / 7 | **Al-KTO#3:** 165 / 12 |

In summary, we successfully stabilized a 2DEG in Si$_3$N$_4$/Al//KTO(110) heterostructures using magnetron sputter deposition through a straightforward and robust procedure. Our findings demonstrate that a Si$_3$N$_4$ cap layer effectively protects the underlying 2DEG located at the Al/KTO interface. Magneto-transport characterization revealed that samples with a nominal Al thickness of 2.5 nm exhibited higher sheet carrier densities but lower mobilities compared to samples with a nominal Al thickness of 1 nm. We propose that the behavior of the sheet carrier density and mobility as a function of temperature in thicker Al layers can be explained by a freeze-out carrier mechanism. In contrast, samples with thinner Al layers likely involve a competition between different mechanisms, including screening of scattering centers, frozen-out carriers, and re-trapping of free carriers into localized states. Using STEM-EELS, we observed the amorphous growth of the Al and Si$_3$N$_4$ layers on top of the crystalline KTO(110). XPS depth profiling revealed that the Al layer in our samples is fully oxidized. Additionally, we demonstrated that the TOF-SIMS technique provides valuable chemical information about the interface, which is crucial for determining the spatial extent of V$_O$'s close to the interface. We found that the V$_O$'s are indicated by variations in the KTaO$_3^-$ and KTaO$^-$ clusters sputtered from the sample during depth profiling, allowing a direct estimation of their distribution. The V$_O$'s, generated by the reducing Al layer, are one of the most significant factors influencing the transport properties in our heterostructures, which can be tuned by adjusting the Al layer thickness. Finally, our structural and electrical characterizations suggest an intuitive correlation: thinner Al layers result in a smaller V$_O$'s extent and lower carrier density, but higher mobility, and vice versa.

## Acknowledgments

This work has been supported by: Comunidad de Madrid (Atracción de Talento grant No. 2022-5A/IND-24230 and MAD2D-CM-UCM3), grant CNS2022-135485 funded by MCIN/AEI/ 10.13039/501100011033 and European Union NextGeneration EU/PRTR, grant from MICINN FEDER PID2021-122980OB-C51, TED2021-129254B-C21 and TED2021-129254B-C22, grants 06/C025-T1 funded by Universidad Nacional de Cuyo, and PIP 11220210100411CO founded by CONICET. E.A.M. acknowledges funding from the European Union (NextGenerationEU), under the Italian Ministry of University and Research (MUR), National Recovery and Resilience Plan (NRRP) grant - TOTEM - CUP E53D23001710006. The transmission electron microscopy was done at CNME/ICTS in Madrid.

## Methods

*Samples fabrication:* All the samples were fabricated through magnetic sputtering deposition with Ar plasma at a substrate temperature of about 250°C. We used KTO substrates from piKEM and SurfaceNet, while the LAO substrates were acquired from Crystec. Before going to the high vacuum chamber, the KTO and LAO substrates were cut in pieces of 2.5x2.5x0.5 mm$^3$ and 2.5x2.5x1.0 mm$^3$, respectively, by using a diamond wire saw. Then, the substrates were rinsed by three cycles of sonication at room temperature in an acetone bath for 5 min, followed by 5 min within an isopropanol bath. The cleaning process is completed inside the vacuum chamber, where the substrates are annealed at T=500°C for 30 min in a pressure better than 10$^{-7}$ mbar. To grow the Al layer, 2-3 W DC sputtering on a metallic Al target under an Ar pressure of 5x10$^{-3}$ mbar was applied. On the other hand, a Si$_3$N$_4$ target was used to grow the cap layer also under an Ar pressure of 5x10$^{-3}$ mbar. To avoid charge effects, due to the insulating nature of the target, the Si$_3$N$_4$ was deposited by 30 W RF sputtering. The deposition rates in these conditions, obtained by thickness calibration with XRR, for the Al and Si$_3$N$_4$ layers were of 1.2 and 0.4 Å/s, respectively. We kept the nominal thickness for the Al layer between 1 and 2.5 nm. For the Si$_3$N$_4$ layer, the nominal thickness ranged from 5 to 20 nm. No differences were observed for different Si$_3$N$_4$ layer thickness, which indicates that, although fundamental for the 2DEG stabilization by protecting the AlO$_x$ interface, it does not play any role in the system properties.

*Magneto-transport measurements:* Once the samples were removed from the vacuum chamber, they were contacted by Al-wire ultrasonic bonding to standard holders in the van der Pauw configuration. Resistance measurements to determine $R_s$ at ambient conditions were made on a 4-tip probe station utilizing a Keithley 2450 Source Measure Unit (SMU) from Tektronix. All $R_s$ and Hall measurements across varying temperatures were performed using a physical properties measurements system (PPMS) device from Quantum Design, allowing temperature control down to 2 K and magnetic fields ranging from -14 T to 14 T.

*STEM-EELS imaging:* The STEM-EELS measurements were performed using a JEOL ARM200cF operated at 200 keV, equipped with a Gatan Quantum EELS system and a spherical aberration corrector.

*TOF-SIMS measurements:* The TOF-SIMS experiments were performed using a TOF.SIMS 5-100, IONTOF GmbH. We applied this technique in depth profile mode as our central interest relied on the composition profiles for the structural characterization of the samples. In this mode, profiles are obtained using two separate sources (dual source): one for conducting the chemical analysis and another for material removal, specifically for surface sputtering. Sputtering was conducted using Cs$^+$ ions at both 250 eV and 500 eV. Ion spectra were captured using pulsed Bi$^{3+}$ ions at 30 keV in every instance. It is worth noting that the sputtering impact of the Bi$^{3+}$ analysis beam is insignificant compared to that of the Cs$^+$ beam. We opted for Cs$^+$ for sample sputtering specifically to record the O content. During these measurements, the internal pressure of the TOF-SIMS chamber ranged between 10$^{-8}$ and 5×10$^{-9}$ mbar. The size of the sputtered area was 250×250 µm$^2$. To avoid any undesired boundary influences, we confined our analysis to a central 100×100 µm$^2$ area within the sputtered crater. After data collection, we designated smaller regions of interest (ROIs) inside the analysis zone to eliminate potential aberrations arising from sample anomalies like holes. After performing the TOF-SIMS measurements, we inspected the sample topography to gauge the depth of each crater using both stylus and optical profilometers. This allowed us to determine average sputter rates for each measurement. At 500 eV Cs the sputter rate resulted almost 10 times higher than at 250 eV due to an increased sputtering yield and a much higher incident current. Once the compositional profiles are obtained, they are normalized to values less or equal to 1, so that the comparison in the spatial distribution of the chemical species is improved.

*XPS measurements:* XPS spectra were collected by using a SPECS Surface Nano Analysis GmbH (SPECS) with a monochromatized Al K$_\alpha$ X-ray source ($hv$ = 1486.7 eV) at room temperature and a pressure better than 10$^{-9}$ mbar. The depth profiling was performed with 1 keV Ar$^+$ sputtering over the entire sample area (2.5 x 2.5 mm$^2$).